\documentclass[conference]{IEEEtran}
\IEEEoverridecommandlockouts
\usepackage{cite}
\usepackage{amsmath,amssymb,amsfonts}
\usepackage{algorithmic}
\usepackage{graphicx}
\usepackage{textcomp}
\usepackage{xcolor}
\usepackage{url}
\usepackage{etoolbox}
\usepackage{subfig}
\usepackage{lmodern}
\usepackage{romannum}
\usepackage{epsfig}
\usepackage{dblfloatfix}

\apptocmd{\sloppy}{\hbadness 10000\relax}{}{}
\usepackage[breaklinks]{hyperref}

\def\BibTeX{{\rm B\kern-.05em{\sc i\kern-.025em b}\kern-.08em
    T\kern-.1667em\lower.7ex\hbox{E}\kern-.125emX}}
\usepackage{amsthm}
\newtheorem*{assumption*}{\assumptionnumber}
\providecommand{\assumptionnumber}{}
\makeatletter

\newenvironment{assumption}[2]
 {%
  \renewcommand{\assumptionnumber}{#1$\mathcal{#2}$}%
  \begin{assumption*}%
  \protected@edef\@currentlabel{#1$\mathcal{#2}$}%
 }
 {%
  \end{assumption*}
 }
\makeatother

\begin{document}

\title{Temporal Assessment of Malicious Behaviors: Application to Turnout Field Data Monitoring  \\

\thanks{This work was supported by the French government and BPI France under the RAILMON project with the funding of the Future Investment Program (Programme d’investissements d’avenir).}
}

\makeatletter
\newcommand{\linebreakand}{%
  \end{@IEEEauthorhalign}
  \hfill\mbox{}\par
  \mbox{}\hfill\begin{@IEEEauthorhalign}
}

\author{\IEEEauthorblockN{Sara Abdellaoui \hspace{-4em}}
\and
\IEEEauthorblockN{Emil Dumitrescu \hspace{-4em}} 
\and
\IEEEauthorblockN{Cédric Escudero \hspace{-4em}}
\and
\IEEEauthorblockN{Eric Zamaï}
\linebreakand
\IEEEauthorblockA{\textit{INSA Lyon, Universite Claude Bernard Lyon 1, Ecole Centrale de Lyon, CNRS,}\\
\textit{Ampère, UMR5005, 69621 Villeurbanne, France}\\
firstname.lastname@insa-lyon.fr}
}
\maketitle

\begin{abstract}
Monitored data collected from railway turnouts are vulnerable to cyberattacks: attackers may either conceal failures or trigger unnecessary maintenance actions. To address this issue, a cyberattack investigation method is proposed based on predictions made from the temporal evolution of the turnout behavior. These predictions are then compared to the field acquired data to detect any discrepancy. This method is illustrated on a collection of real-life data. 
\end{abstract}

\begin{IEEEkeywords}
turnout, cyberattack, forecasting, time series, railway

\end{IEEEkeywords}
\vspace{-4mm}
\section{Introduction}\label{Section 1}

Cyber Physical Systems (CPS) represent physical processes integrated with computational components \cite{lee_cyber_2008}. Their distributed communicating nature makes them vulnerable to cyberattacks
\cite{escudero_process-aware_2018}. The security of CPS has emerged as a complex problem, after discovering the Stuxnet malware \cite{
franck_ics_2018} that targeted the Iranian industrial control system. 

Among CPS, railways systems are subject to critical safety requirements regarding the security of people and equipment \cite{leyden_polish_2008, tabak_ransomware_2021}, as well as the availability, and the accessibility \cite{
roth_cyberpartisans_2022}.     
According to the European Union Agency for Cybersecurity (ENISA) report \cite{european_union_agency_for_cybersecurity_enisa_2023}, the railway sector faces intensive cyberthreats: 21\% of the total observed incidents between January 2021 to October 2022. The primary goals of these incidents were to disrupt operations and to gain financial benefits \cite{european_union_agency_for_cybersecurity_enisa_2023}.


This article proposes a method to identify cyberthreats targeting railway systems, specially turnout systems, as presented in Fig.~\ref{Turnout}. Turnouts are essential elements of railways that facilitate the change of trains' direction 
by executing switch operations. 
They are subject to critical requirements in terms of security and reliability. 
Maintenance operators are responsible for keeping the turnout running.
To do so, they need to decide about the health state of the turnout, based on the railway monitoring system information. However, this system is vulnerable to cyberattacks. This paper focuses on cyberattacks causing the occurrence of False Positives (FP) or False Negatives (FN) anomaly reports \cite{abdellaoui_cyber_2023}, leading maintenance operators to take erroneous decisions: either perform an unnecessary maintenance action (FP) or worst, doing nothing while the turnout is malfunctioning by hiding existing failures (FN).  



This article takes over the issues addressed in \cite{abdellaoui_cyber_2023}: help maintenance operators by detecting cyberattacks that aim at concealing failures or causing unnecessary maintenance actions. The proposed approach is based on a collection of monitoring data that need to be assessed in order to extract insights about possible cyberthreats. The shapes of the current consumption curves during switch operations were analyzed and 2 shapes types were taken into consideration, normal fault-free shapes and failure shapes. This information was projected on the life cycle of the turnout according to time aging and operation aging criteria in order to compute a cyberthreat likelihood for each current curve observed. Maintenance operators use the estimated likelihood to assess the authenticity of each piece of collected field data. Despite the context that this approach gives to each new piece of field data with regard to the life cycle of the turnout, it only considers data individually, regardless of its context of evolution.

To overcome this drawback, this work attempts to associate a contextual meaning for every piece of field data, according to the temporal evolution of the field data. 
This approach enables the study of the temporal behavior possibly encompassing the natural aging of a turnout, which is instrumental in identifying cyberattacks that aim to disrupt the turnout behavior.

The rest of the paper is structured as follows. Section \ref{Section 2} provides short state-of-the-art methods for cyberattacks detection and temporal data prediction 
Section \ref{Section 3} introduces the turnout architecture and necessary assumptions are made to develop the proposed method. Section \ref{Section 4} details the development of the detection method, including 1) building a forecasting model by analyzing temporal turnout behavior, 2) predicting an expected turnout behavior curve using the forecasting model, and comparing it to the acquired curve to identify differences, 3) evaluating the likelihood of cyberthreats based on the nature of the acquired curve differences. Section \ref{Section 5} illustrates the effectiveness of the proposed method. Finally, Section \ref{Section 6} concludes the results presented in this article and provides future research directions. 

\begin{figure}[ht]
\vspace{-2mm}
\centerline{\includegraphics[scale=0.4]{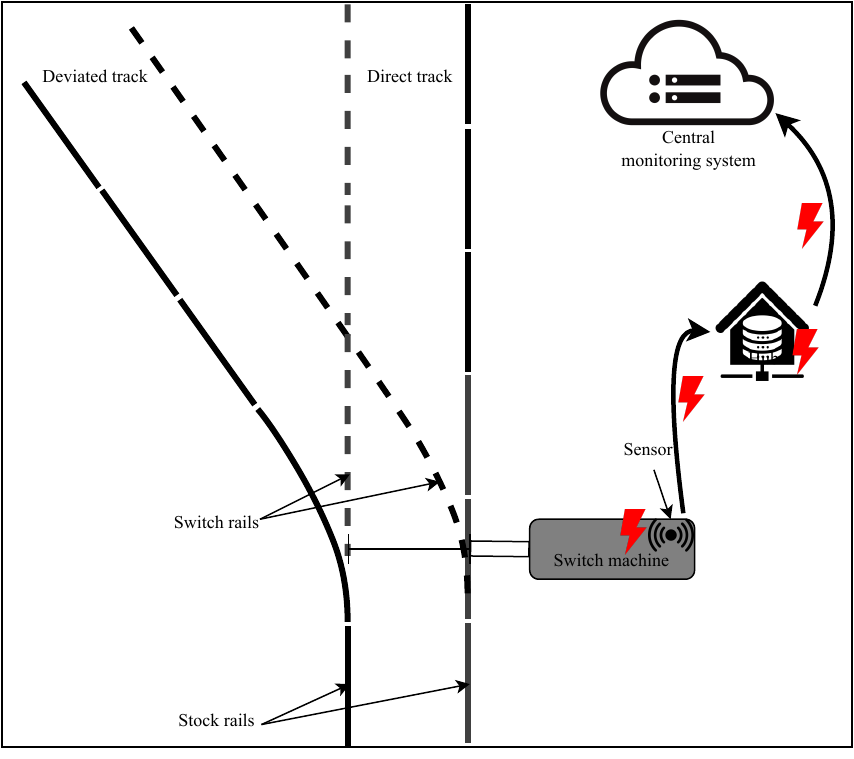}}
\vspace{-2mm}
\caption{Turnout.}
\vspace{-3mm}
\label{Turnout}
\vspace{-3mm}
\end{figure}
\section{Related work}\label{Section 2}

Detection and prediction are the two main approaches dealing with cyberattacks \cite{pino_towards_2014}. According to \cite{pino_towards_2014}, detection is defined as the identification of malicious behaviors that have already happened or initiated and prediction is defined as the characterization of malicious behaviors that are likely to happen in the future.

For detection methods, as mentioned in \cite{abdellaoui_cyber_2023}, the typical Intrusion Detection Systems (IDS) employed for detecting cyberattacks involve either identifying known attacks (Signature-based IDS) or identifying attacks that deviate the system from its predefined behaviors (Anomaly-based IDS) \cite{escudero_process-aware_2018}. As a result, these approaches are unable of detecting threats that hide behind an expected behavior of a system. 

Regarding prediction methods, 4 use cases of prediction in cybersecurity are defined in \cite{husak_survey_2019}: (\romannum{1}) attack projection (predict the next move of an attacker after observing an attack), (\romannum{2}) attack intention recognition (predict the ultimate goal of an attacker), (\romannum{3}) attack prediction (predict the type, when and where an attack will occur), and (\romannum{4}) network security situation forecasting (predict the system overall situation and not just a specific attack).  


\begin{figure*}[ht]
\vspace{-7mm}
\centerline{\includegraphics[scale=1.3]{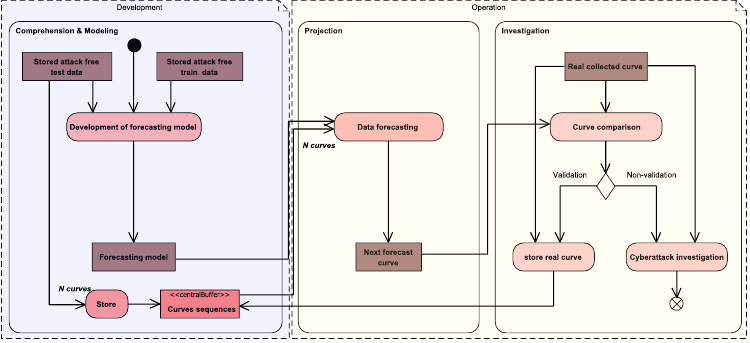}}
\vspace{-2mm}
\caption{Framework proposed.}
\vspace{-5mm}
\label{framework}
\vspace{-2mm}
\end{figure*}

Since the cyberattack types considered in this work are related to the turnout expected behavior, it is more appropriate to rely on use case (\romannum{4}) methods where the notion of ``Situational Awareness" (SA) is applied. Reference \cite{endsley_situation_1988} defines the SA of a system as \emph{the perception of the elements in the environment within a volume of time and space, the comprehension of their meaning, and the projection of their status in the near future}. Then later in \cite{endsley_toward_1995}, the author elaborates on the three levels of the SA: (\romannum{1}), the perception of elements in the environment by collecting and processing the related data, (\romannum{2}), the comprehension of current situations by integrating data from level 1 with pertinent goals and determine their impacts, (\romannum{3}), the projection of future actions through knowledge of data and comprehension of situations.
In 2014, \cite{franke_cyber_2014} introduced ``Cyber Situational Awareness" (CSA) as a subset of the SA that focuses on the ``cyber" environment issues which are combined with other environment's information and data to achieve a high level of comprehension. To achieve CSA, authors in \cite{jiang_systematic_2022} define also 3 levels that aligned with SA levels: (\romannum{1}), perceive the present condition of assets and cyberthreat scenarios (perception), (\romannum{2}), comprehend the significance of the cyberthreat situation and evaluating its impact (comprehension), (\romannum{3}), foresee future action or threat possibilities (projection).

Turnout data have the particularity of being time series data, which are  observations related to switch operations, collected sequentially over time. Authors in \cite{box_time_1978} define the notion of ``time series analysis" as the techniques for analyzing the dependencies between time series data sequences.
The combination between time series analysis and SA are widely used in identifying anomalies and faults, by discovering unusual subsequences relying on the SA three levels: perception, comprehension and projection. 
In \cite{keogh_finding_2002}, authors propose a technique to detect surprising patterns in time series data according to the frequency of their occurrence. Using a reference collection of normal data, the proposed algorithm, named TARZAN, encodes the frequency of all observed patterns and then predicts the expected frequencies of patterns in new data set. 
To detect the surprising patterns, frequencies of all patterns are measured, whereby surprising patterns are those whose frequencies differ significantly from those predicted by the algorithm. The proposed method is proven to be effective in detecting anomalies when compared to two other methods proposed by \cite{shahabi_tsa-tree_2000}, and \cite{dasgupta_novelty_1996}.
Later in \cite{keogh_hot_2005}, authors introduce time series discords as anomaly detectors. The main strength of this approach is that only one parameter is required, which is the length of the subsequence that covers typical usual data in order to detect those that are not. The case of turnout data and the cyberattacks considered in this work is different, since the interest is only on data that are considered usual and expected. 
In the railway sector, \cite{garcia_time_2010} proposes a fault detection algorithm  based on  sequences of turnouts' current curves that represent fault free normal behaviors, A Vector Auto-Regressive Moving-Average (VARMA) is then used to forecast the next curve. If the measured curve deviates from the predicted one, which is obtained based on fault free data, a fault presence is indicated. 

Regarding cybersecurity, based on the state of the art, databases containing attack data or attack features are needed in order to predict the occurrence of attacks in the future. For instance, authors in \cite{werner_time_2017} propose a method using Auto-Regressive Integrated Moving Average (ARIMA) time series forecasting model to predict the intensity of cyberattacks on a future date based on the count of past attack incidents. In \cite{werner_forecasting_2018}, an approach to predict future cyberattacks is proposed. This approach is based on extracting features from past attack data that show correlation to future incident counts and then use ARIMA models and Bayesian Networks (BN) to predict the intensity of future attacks with taking into consideration the attacks' intraday trends. 
In \cite{liu_cloudy_2015}, an approach is proposed to forecast an organization networks' breaches using time series data representing characteristics of previous attacks to train a Random Forest classifier. 

To summarize, existing cyberattacks detection and prediction methods are based on previously observed cyberattacks. As a result, the lack of this information represents a challenge, especially when the only available information to understand the switch operation is the power curve with no prior knowledge of cyberattacks. 

The current article introduces a cyberthreat assessment approach based on prediction and detection. 
This is achieved by analyzing and learning sequences of field data. A notion of expected turnout behavior is defined and implemented. The cyberattack detection is achieved according to the differences observed between the field data collected and its expected behavior.

\section{Context and problem statement}\label{Section 3}


\subsection{Context and system architecture}\label{3-A}

The monitoring framework is shown in Fig.~\ref{Turnout}. The railway turnout system executes switch movements that are captured by sensors. These sensors measure the current and the voltage in the electrical motor displacing the switch. Then, the power consumption of the motor is computed; referred thereafter as current, voltage, and power data.
Subsequently, the data are transmitted to the Central Monitoring System (CMS) through data hubs.


The operational principle of turnout systems states that the current data provide information about the status of the control circuit for switch machines, while the power data reflect the operational condition and deterioration of the mechanical parts of the turnout \cite{zhang_fault_2022}. Consequently, since the considered cyberattacks are related to the operational conditions of the turnout, the power curves describing the turnout operations are chosen to be studied. 



\subsection{Assumptions}\label{3-B}

The proposed method relies on five assumptions.
\vspace{-2mm}
\begin{assumption}{1}{}\label{H1}
\emph{Attackers possess abundant resources to gain comprehensive knowledge of the railway infrastructure, excluding information regarding the turnout life cycle.}
\end{assumption}
\vspace{-4mm}
\begin{assumption}{2}{}\label{H2}
\emph{Attackers have the means to manipulate field data in order to change the shape of a power curve.}
\end{assumption}
\vspace{-4mm}
\begin{assumption}{3}{}\label{H3}
\emph{Compromised field data is never spurious, they mirror expected turnout behavior, representing either normal or failure scenarios.}
\end{assumption}
\vspace{-4mm}
\begin{assumption}{4}{}\label{H4}
\emph{Datasets used in this study are securely obtained and considered to be not compromised.}
 \end{assumption}
\vspace{-4mm}
 \begin{assumption}{5}{}\label{H5}
\emph{No particular discrimination criteria are taken into account to differentiate between turnouts}
 \end{assumption}

\subsection{Problem Statement}\label{3-C}
According to the context presented above, field data can be compromised by attackers as illustrated in Fig.~\ref{Turnout}: physical sensors' reading can be overridden by spurious data, in addition to the fact that communications between sensors and the CMS can be intercepted. 


The problem addressed in this work is the evaluation of  the authenticity of a turnout field power curve by situating it within its operation context: given a sequence of power curves witnessing a long evolution of the turnout health, compare each piece of field data to its expected shape in order to assess the likelihood of a cyberthreat. 




\vspace{-2mm}
\section{Detection Method Development}\label{Section 4}

\subsection{Overview}\label{4-A}

The cyberattack assessment is achieved by predicting the next expected field data, according to the data history already observed. If the field data is different from the predicted data, further investigations are conducted. 
This approach is outlined in Fig.~\ref{framework} and follows three stages derived from SA levels:
\begin{itemize}
    \item \textit{Comprehension \& Modeling}: Development of a forecasting model 
    \item \textit{Projection}: Prediction of a switch operation curve 
    \item \textit{Investigation}: Comparison between the predicted and collected curves and decision regarding the malicious suspicion 
\end{itemize}

These stages are implemented in two distinct phases, as illustrated in Fig.~\ref{framework}: the development phase and the operation phase.


\vspace{-1mm}
\subsection{The Development phase}\label{4-B}


Predicting the behavior of a turnout requires the development of a forecasting model. An initial collection of data sets is used, and in accordance with assumption \ref{H4}, these data sets are considered non-compromised.

The data used in this phase represent field data acquired during the turnout useful life and capture fault-free behaviors, progressive pre-fault anomalies and aging behaviors.

 According to \cite{han_review_2021}, prediction methods over data sequences are divided into discriminatory methods, where the target prediction is based on learning from observed data, and generative methods where the prediction is based on the distribution of observations and targets. The goal of this work is to predict the behavior of a turnout based on previous observations, as a result, discriminatory methods are considered. In \cite{han_review_2021}, a comparison was made between different discriminatory methods, such us Recurrent Neural Networks (RNN), Convolutional Neural Network (CNN), Long-Short Term Memory (LSTM), Generative Adversarial Nets (GANs), Deep Belief Network (DBN) and a hybrid model of sparse Auto-Encoder (AE) with LSTM. Experiments and real world tests show that LSTM and hybrid model perform better. Hence, the LSTM technique is chosen to develop a forecasting model. Its ability to generalize from both long and short term sequences makes it a good candidate for our solution. 

Fig.~\ref{LSTM} illustrates the specific data preparation operation applied in order to train the LSTM model: a sequence of N curves is used to predict the $N+1^{th}$. For a total number $M$ of switch operation curves in train data, the size of the training data is $M-N$. The same preparation operation is applied to the test data. 
The resulting  trained and validated model represents the forecasting model used throughout this framework. 


\begin{figure}[!b]
\centerline{\includegraphics[scale=0.8]{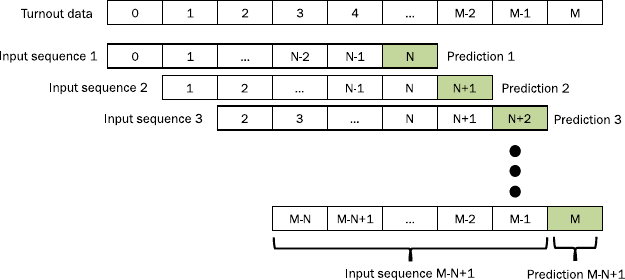}}
\vspace{-2mm}
\caption{Data preparation for the LSTM learning.}
\label{LSTM}
\end{figure}

\subsection{The Operation phase}\label{4-C}

The operation phase refers to the real life operation of the implemented forecasting model, as shown in Fig.~\ref{framework}.  
As highlighted in Fig.~\ref{LSTM}, the forecasting model needs an initial sequence of $N$ curves, in order to predict the $N+1^{th}$. This sequence evolves over time: new field data is added at each turnout operation, while old data are discarded. 
At the very beginning of the prediction process, the initial sequence fed to the predictive model is constructed from the test data. 

\subsubsection{Projection}\label{4-C-1}

In this step, 
the forecasting model predicts the next power curve expected from a sequence of previously observed power curves. 

\subsubsection{Curve Comparison}\label{4-C-2}

The prediction is followed by an investigation:
the predicted curve is compared to the acquired field curve in order to spot possible differences between the reality and the expected behavior. 
As this comparison is performed on power curves, a suitable metric is required. 
The Euclidean distance and Dynamic Time Warping (DTW) are commonly used methods for this purpose. DTW is a powerful similarity measure algorithm that can capture time axis warping or stretching, as it allows for one-to-many point comparisons. However, DTW may struggle to capture significant irregularities between curves that have the same shape but differ in phase shift. To overcome this limitation, in addition to the DTW algorithm, and since the compared curves have the same length, the Euclidean distance could also be computed to ensure a robust and comprehensive comparison. 
As a result, the comparison is carried out using two criteria. In order to define their corresponding thresholds, the forecasting model was used on test data to help identify the value from which two curves are deemed to be different.

Based on DTW and Euclidean distances thresholds, a field curve may either be considered similar to the predicted curve (validated) or different (non-validated). If the field curve is found to be similar to the expected curve, it will be stored in ``Curves sequences" storage unit to use it in the next input sequence for the forecasting model to ensure temporal continuity. Otherwise, further investigations are needed to assess the likelihood of a cyberthreat.  
\begin{figure}[!b]
\vspace{-5mm}
\centerline{\includegraphics[scale=0.7]{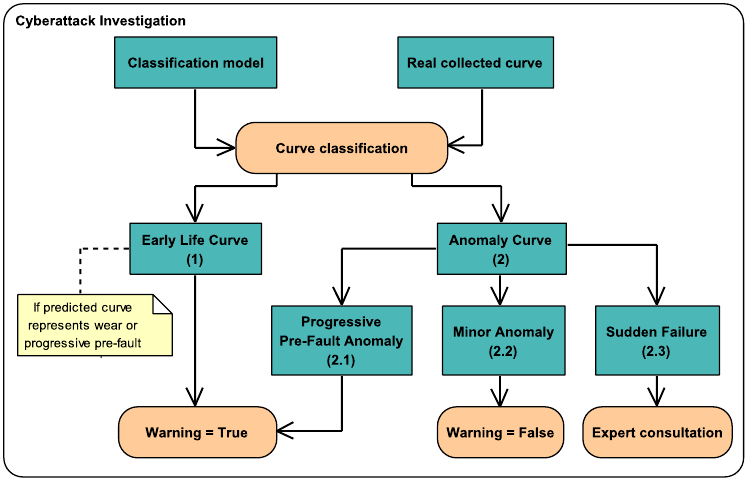}}
\vspace{-2mm}
\caption{Cyberattack investigation process.}
\vspace{-7mm}
\label{Cyber_inv}
\end{figure}

\subsubsection{Cyberattack Investigation}\label{4-C-3}


The decision-making process is described in Fig.~\ref{Cyber_inv}. It relies on a pre-existing classification model \cite{abdellaoui_cyber_2023}. 
Two possible scenarios are considered: 

\begin{figure*}[!b]
\vspace{-3mm}
\centering{\includegraphics[scale=0.38]{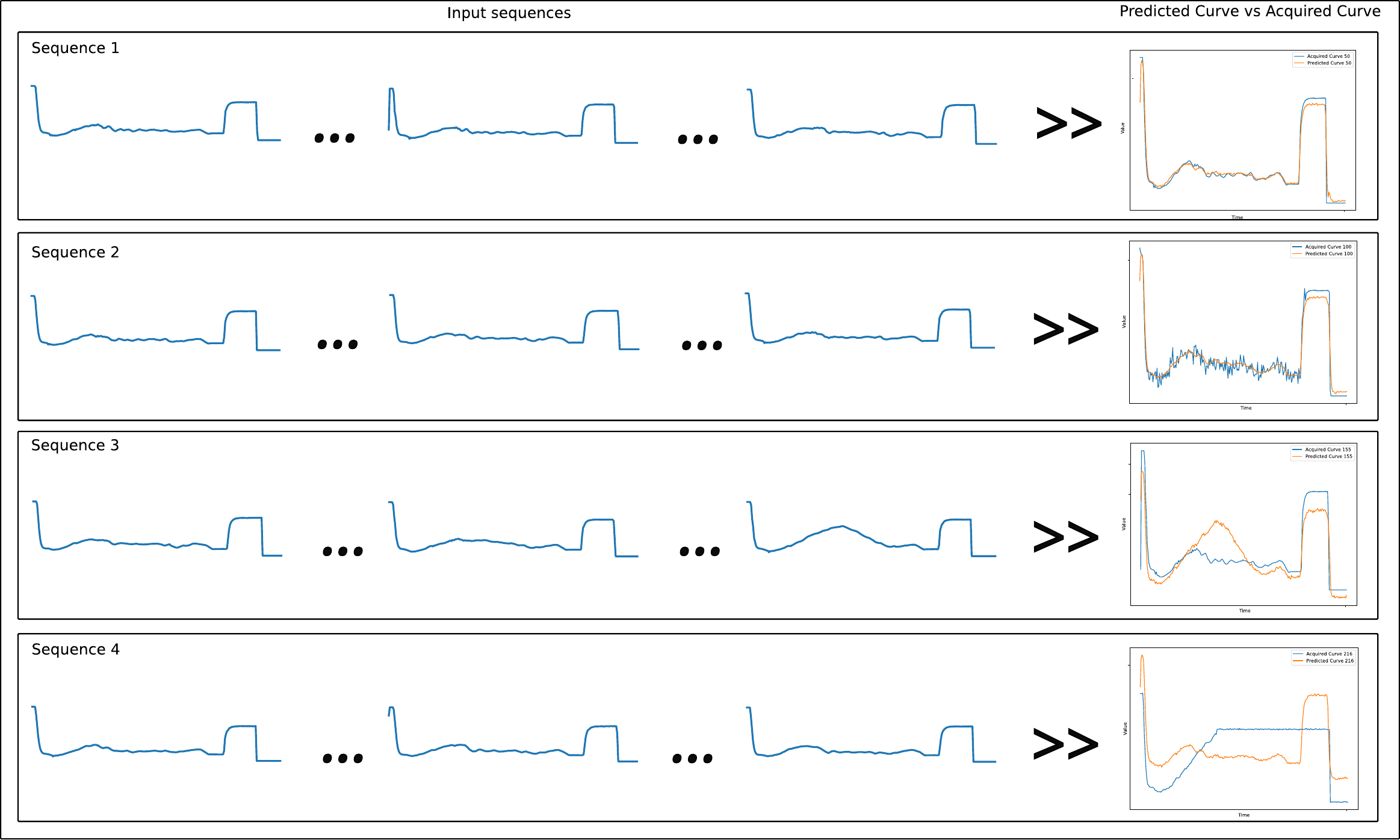}}
\vspace{-1mm}
\caption{Input sequences \& predictions.}
\vspace{-7mm}
\label{input_pred}
\end{figure*}

\begin{itemize}
    \item The field curve represents a ``normal", early life behavior: this implies that the predicted curve represents  either  a progressive pre-fault anomaly or an end of life curve, based on a sequence which establishes a progression towards an anomaly. Hence, receiving an early life behavior is unexpected and suspicious. 
    
    \item The field curve represents an anomaly: this scenario involves 3 cases. (\romannum{1}), the anomaly corresponds to a progressive pre-fault anomaly. Therefore, it is deemed suspicious to have this curve without any prior indication of progressive anomalies. (\romannum{2}), the anomaly curve represents a minor anomaly that resolves naturally without the need for maintenance intervention, in this case no suspicion is raised. (\romannum{3}), the anomaly curve reflects a sudden failure. Since the forecasting model is unable to predict behaviors that manifest suddenly, additional expertise is required to make a decision. However, this information falls outside the scope of this paper. 
\end{itemize}

\vspace{-3mm}
\section{Case study}\label{Section 5}

A turnout with non-compromised  data, i.e., satisfying Assumption \ref{H4} describing almost 1000 switch operation is chosen. The selected data contain both ``normal" early life behaviors and behaviors reflecting aging and progressive pre-fault anomalies. 

The LSTM hyperparameters are determined through multiple experiments:  the mean square error (MSE) loss function was found to be efficient for comparing curves point by point. The optimal number of LSTM layers was determined by profiling models with one, two, and three layers. The single-layer model achieves the best result. The LSTM input window size yielding the best learning performance is $N = 50$. 
The Adam optimizer \cite{kingma_adam_2014} is chosen for its extensively proven effectiveness \cite{chang_electricity_2019}.

Threshold values need to be estimated for the Euclidean distance and the DTW distance,  in order to define the separation between accepted and rejected curves. For this purpose, the LSTM model is applied on 
test data in order to determine the maximum values of the two distance metrics between two curves that are judged to be sufficiently close. 

Fig.~\ref{input_pred} illustrates examples of the comparison between predicted curves and the corresponding field data acquired. Each predicted curve is obtained from a 50-curves sequence featuring a possible evolution in the turnout operation. 
Sequences 1 and 2 represent early life turnout behaviors. As a result, the forecasting model also predicts early life curves. It can be seen that the second prediction 
features a mismatch with respect to the field data. This raises a cyberattack suspicion, since the acquired curve indicates a sudden aging scenario ((2)-(2.1) in Fig.~\ref{Cyber_inv}). Sequence~3 features a progressive pre-fault anomaly, which results in the prediction of a curve representing the natural expected evolution for this kind of anomaly. Yet the field data represent an early life curve, which is suspicious ((1) in Fig.~\ref{Cyber_inv}). Sequence 4 features the sudden occurrence of a failure according to the field data. Yet, this failure should have been preceded by a progressive evolution which is absent, which is why this scenario is considered suspicious ((2)-(2.1) in Fig.~\ref{Cyber_inv}).

These results demonstrate the ability of the forecasting model to assess the authenticity of field curves with respect to the temporal operation context of the monitored turnout. 
One main limitation encountered so far is  the inability to capture slowly progressive aging scenarios: it seems challenging for LSTM to capture the spatial variation characteristics of noise \cite{tiwari_auto-encoder_2022}. 
\section{Conclusion}\label{Section 6}
A cyberattack detection method based on analyzing the turnout temporal behavior has been presented and illustrate on real-life examples. 

The proposed method relies on predicting a power curves related to the expected behavior of the turnout, with respect to its observed past evolution. 
Differences between, the field data and the predicted data lead to a cyberattack investigation process. 

Future developments shall address the current limitation of LSTM prediction for slowly aging curves. Auto-encoder based models with LSTM, 
seem promising in overcoming this limitation. The current approach for cyberthreat assessment can be enhanced by taking into account the life cycle of a turnout. A unified approach encompassing \cite{abdellaoui_cyber_2023} can be promising. 


Finally, it is essential yet challenging to generalize the detection method for a heterogeneous collection railways turnouts:
they always exhibit  unique behaviors following their location, temperature, weather conditions. 
Future research should identify discriminant parameters between turnouts and explore turnout behaviors according to these parameters. 


\




\bibliographystyle{IEEEtran}
\vspace{-8mm}
\bibliography{IEEEabrv, Biblio_Zotero}

\end{document}